\documentclass[12pt]{article}

\usepackage{a4}
\usepackage{epsfig}

\newlength{\abstwidth}
\def\be{\begin{equation}} 
\def\ee{\end{equation}} 
\def\bqa{\begin{eqnarray}} 
\def\eqa{\end{eqnarray}} 
\def\lsim{\roughly<}

\begin{document}

\setcounter{section}{0}

\renewcommand{\theequation}{%
\mbox{\arabic{section}.\arabic{equation}}}

\def\lsim{\mathrel{\rlap{\lower4pt\hbox{\hskip1pt$\sim$}}
    \raise1pt\hbox{$<$}}}         
\def\gsim{\mathrel{\rlap{\lower4pt\hbox{\hskip1pt$\sim$}}
    \raise1pt\hbox{$>$}}}         

\pagestyle{empty}

\begin{flushright}
BI-TP 2002/21\\
~ \\

\end{flushright}

\vspace{\fill}

\begin{center}
{\Large\bf Deep inelastic scattering and ``elastic'' diffraction}
\\[1.8ex]
{\bf Masaaki Kuroda} \\[1.2mm]
Institute of Physics, Meiji-Gakuin University \\[1.2mm]
Yokohama 244, Japan \\[1.2mm]
and \\[1.5ex]
{\bf Dieter Schildknecht} \\[1.2mm]  
Fakult\"{a}t f\"{u}r Physik, Universit\"{a}t Bielefeld \\[1.2mm] 
D-33501 Bielefeld, Germany \\[1.5ex]
\end{center}

\vspace{\fill}

\begin{center}
{\bf Abstract}\\[2ex]
\begin{minipage}{\abstwidth}
We examine the total cross section of virtual 
photons on protons, $\sigma_{\gamma^* p}(W^2,Q^2)$,
at low $x \cong Q^2/W^2 \ll 1$ and its connection with 
``elastic'' diffractive production $\gamma^*_{T,L}p \rightarrow X^{J=1}_{T,L} p$
in the two-gluon exchange dynamics for the virtual forward Compton scattering
amplitude. Solely based on the generic structure of two-gluon exchange, 
we establish that the cross section
is described by the (imaginary part of the) amplitude for forward scattering 
of $q \bar q$ vector states, $(q \bar q)^{J=1}_{T,L} p \rightarrow (q \bar q)^
{J=1}_{T,L} p$. The generalized vector dominance/color dipole picture 
(GVD/CDP) is accordingly established to only rest on the two-gluon-exchange 
generic structure. This is explicitly seen by the sum rules that allow one to 
directly relate the total cross section to the cross section for elastic 
diffractive forward 
production, $\gamma^*_{T,L} p\rightarrow (q \bar q)^{J=1}_{T,L} p$, of 
vector states. 

\end{minipage}
\end{center}

\vspace{\fill}
\noindent


\noindent

\clearpage
\pagestyle{plain}
\setcounter{page}{1}

\baselineskip 20pt

\section{Introduction}

A widely accepted picture of deep inelastic electron scattering on nucleons
at low values of $x \cong Q^2/W^2\ll 1$, in terms of the virtual forward Compton 
amplitude, is based on the two-gluon exchange dynamical mechanism \cite{a}
depicted in fig.~1. The two-gluon exchange mechanism was evaluated \cite{b} 
in momentum space and its representation \cite{b} in transverse position space 
\cite{c}
became known as the color-dipole approach \cite{b}:
taking into account the low-$x$ kinematics, the photon in fig.~1 virtually 
dissociates into a $q \bar q$ color dipole that subsequently undergoes diffractive
forward scattering on the proton. 

\begin{figure}[htbp]\centering
\epsfysize=5cm
\centering{\epsffile{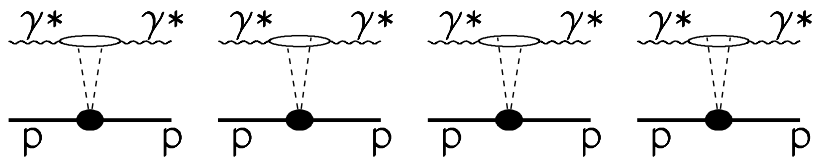}}
\caption{The forward Compton amplitude.}
\label{fig1}
\end{figure}

\bigskip\noindent
In the present work, we examine the question of which spins of the $q \bar q$ 
system contribute to the forward scattering process $(q \bar q) p \rightarrow 
(q \bar q)p$ in the Compton amplitude. It comes without surprise that we find 
that the process exclusively proceeds via the forward scattering of $J=1$ 
(i.e. vector) $q \bar q$ states, $(q \bar q)^{J=1}$. Our result is only based on 
the 
generic structure of the two-gluon exchange in fig.~1. Obviously, $J=1$ is a 
consequence of the $\gamma^* q \bar q$ transition that only allows for 
interactions of $J=1$ states, $(q \bar q)^{J=1} p \rightarrow (q \bar q)^{J=1} p$.
The relevance of only $(q \bar q)^{J=1}$ states is most transparently expressed
in terms of a sum rule that relates the total virtual photoabsorption cross 
section to diffractive forward production (compare fig.~2) of vector states, 
$\gamma^*_{T,L}p\rightarrow (q \bar q)^{J=1}_{T,L} p$.
The sum rule to be given in the present paper  
coincides with the one obtained in ref. \cite{d} under more 
restrictive assumptions \cite{e} on the $q \bar q$ interaction with the proton 
corresponding 
to the lower vertices in fig.~1. In the present work, no specific
ansatz for the gluon-gluon-$pp$ interaction is introduced, and, in addition,  
the structure of the previously adopted ansatz \cite{e} 
is recognized as a realization,
without much loss of generality, of the underlying 
$(q \bar q)^{J=1} p \rightarrow (q \bar q)^{J=1} p$ interaction.

\begin{figure}[htbp]\centering
\epsfysize=5cm
\centering{\epsffile{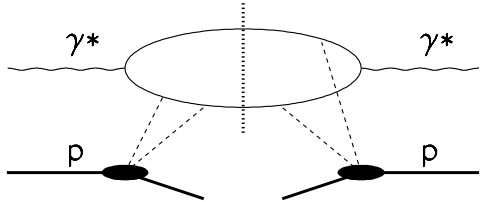}}
\caption{One of the 16 diagrams for diffractive production.  
The vertical line indicates the unitarity cut corresponding to the
diffractively produced final states, $(q\bar q)^J$.  
Production of (discrete or continuum) vector states corresponds to 
$(q\bar q)^J$ production with $J=1$.}
\label{fig2}
\end{figure}

The emerging picture of deep inelastic scattering at low $x$ 
coincides with the 
one of generalized vector dominance (GVD) \cite{f,g}
\footnote{Compare also the review on GVD in \cite{Shaw}}
from the pre-QCD era;
generalized vector dominance is obtained as a consequence of 
the two-gluon exchange generic 
structure from QCD. As previously stressed \cite{h}, 
it is precisely the gauge-theory structure underlying 
the two-gluon interaction 
with its change in sign between the different contributions in fig.~1 that 
is responsible for the consistency of GVD. This structure, 
in terms of cancellations between diagonal and off-diagonal 
transitions, was anticipated\cite{h,i} by off-diagonal GVD \cite{g}.

The present work clarifies the connection between the total cross section and 
diffractive production as observed \cite{j} at HERA:
the total cross section, $\sigma_{\gamma^*_{T,L}p} (W^2 , Q^2)$, is 
quantitatively related to the amplitudes of ``elastic'' diffractive production, 
$\gamma^*_{T,L}p \rightarrow (q \bar q)^{J=1}_{T,L}p$, via the 
above-mentioned sum rule. ``Inelastic'' diffraction,  
$\gamma^*_{T,L}p \rightarrow (q \bar q)^{J\not= 1}p$, (corresponding
to diffraction dissociation in hadron-hadron interactions) constitutes an 
additional important contribution to diffractive production, in particular for 
high invariant masses of the states produced. Inelastic diffraction is 
irrelevant, however, with respect to 
the total cross section. It is (obviously) only the elastic 
component of diffractive 
production that enters the forward Compton amplitude. 

Fits to the experimental data for $\sigma_{\gamma^* p} (W^2 , Q^2)$ or, 
equivalently $F_2 (x, Q^2)$, were frequently based \cite{k,l,m}
on an ansatz for the color-dipole cross section in transverse position space that
did not incorporate the restriction to $(q \bar q)^{J=1}$ states. Since only 
$(q \bar q)^{J=1}$ states contribute to the scattering, while all others,
$(q \bar q)^{J\not= 1}$, yield vanishing contributions, the latter ones 
implicitly remained 
undetermined in the fits to $\sigma_{\gamma^* p} (W^2 , Q^2)$. This lack of 
restrictions on the amplitudes for $(q \bar q)^{J = 1}p \rightarrow 
(q \bar q)^{J\not= 1} p$ is presumably the reason for widely differing results
\cite{n}
on the dipole cross section that were extracted from the fits. Dropping the 
redundant $J\not = 1$ terms 
in fits to the total cross section, $\sigma_{\gamma^*_{T,L}p}(W^2 , Q^2)$, right 
from the beginning,
most likely will improve the uniqueness of the 
extracted dipole cross sections. We note that the ansatz of the 
generalized vector dominance/color dipole picture
(GVD/CDP) \cite{e} 
incorporates the restriction to $(q \bar q)^{J=1}$ states. 

In section 2, we use the conventional formalism of the color-dipole model in 
transverse position space, in order to show that $\sigma_{\gamma^* p} (W^2 , Q^2)$
is determined by the $(q \bar q)^{J= 1}p \rightarrow (q \bar q)^{J= 1} p$
forward scattering amplitude, i.e. by the color-dipole cross section for 
$J=1$ color-dipoles originating from the $\gamma^* \rightarrow q \bar q$ 
dissociation. We show how the redundant amplitudes from scattering into
$(q \bar q)^{J\not= 1}$ states can be eliminated 
right at the beginning
by a slight refinement in the 
representation of the total cross section in transverse position space. 

In section 3, we use the momentum-space representation in order to derive  the 
sum rules that determine the longitudinal and transverse photoabsorption 
cross sections,
$\sigma_{\gamma^*_{T,L}p}(W^2 , Q^2)$, as integrals over the 
mass spectra of the diffractively produced vector states. 

In section 4, we elaborate on the connection between the dipole cross 
section and the gluon structure function. 

In section 5, we remind the reader of the ansatz \cite{e} for the 
color-dipole cross section 
used in the fits to the total cross section in the GVD/CDP.
 
In section 6, we stress the duality relation between the 
description of low-$x$ deep inelastic scattering in  terms of
scattering of $(q\bar q)$ vector states,
$(q\bar q)^{J=1} p \to (q\bar q)^{J=1}p$, in 
the GVD/CDP, and a description in terms of $\gamma^*$gluon
scattering, $\gamma^* g\to \gamma^* g$ employing the notion of the
gluon structure function.
The  gluon structure function of the GVD/CDP multiplied by
$\alpha_s(Q^2)$ becomes a function of a single variable 
$1/W^2=x/Q^2$.  For any fixed value of $Q^2$, conventional
evolution of the structure function breaks down for
$x<x_0(Q^2)$ where $x_0(Q^2)$ is calculable in the GDV/CDP.
For values of $x<x_0(Q^2)$ "saturation" occurs in the sense  of
$\sigma_{\gamma^*p}(W^2,Q^2)/\sigma_{\gamma p}(W^2) \to 1$.
Conversely, for any fixed $x\ll 0.1$, the conventional evolution
holds for the gluon structure functions, provided
$Q^2$ is sufficiently large.

A few concluding remarks will be given in section 7.

\section{Two-gluon exchange in transverse-position-\\
         space representation and $J=1$}

The transverse-position-space representation \cite{c,b}
for the total photoabsorption cross section,
\be
\sigma_{\gamma^*_{T,L} p} (W^2 , Q^2 ) = \int dz \int d^2 r_\bot \left| 
\psi_{T,L} (r_\bot , z (1-z) , Q^2 ) \right|^2 
\sigma_{(q \bar q )p} (\vec r_\bot ,  W^2 ), 
\label{(27)}
\ee 
and for diffractive forward production,
\begin{eqnarray}
\left.\frac{d\sigma_{\gamma^*_{T,L} p \rightarrow Xp} (W^2 , Q^2)}{dt}  
\right|_{t=0}
  &=& \frac{1}{16\pi} \int dz \int d^2 r_\bot \left| \psi_{T,L} 
(r_\bot , z(1-z) , 
    Q^2 ) \right|^2  \nonumber \\ 
& &~~~~~~~~~  \sigma^2_{(q \bar q )p} (\vec r_\bot ,  W^2 ) ,
\label{(28)}
\end{eqnarray}
conveniently summarizes the result of the $x\cong Q^2/W^2 \rightarrow 0$ 
analysis of the two-gluon-exchange dynamics. 
As 
intuitively suggested by the underlying $s$-channel point of view, we use the 
center-of-mass energy $W$ of the photon-proton (equivalently, of the 
$q \bar q$-proton)
system as a variable in the dipole cross section. The assumed dependence on 
$W^2$ proved useful in the representation of the experimental data \cite{e}, as 
it automatically, and naturally, via $W^2 \simeq Q^2/x$, induces scaling 
violations for the structure function $F_2 (x, Q^2) \cong (Q^2/4\pi^2\alpha)
\sigma_{\gamma^*p}(W^2,Q^2)$ as observed experimentally\cite{devenish}.
Frequently \cite{l} $x$ replaces $W^2$ 
in (\ref{(27)}), requiring a revision \cite{m} of the original ansatz to 
incorporate scaling violations. 

The wave function $\psi_{T,L}(\vec r_\bot, z (1-z), Q^2)$ describing the 
$\gamma^* q \bar q$ fluctuation of the virtual photon in (\ref{(27)}) and 
(\ref{(28)}) has the well-known form \cite{b} 
\be
\left| \psi_{T} (r_\bot , z (1-z), Q^2 ) \right|^2 = \frac{6 \alpha}{(2\pi)^2} 
\sum_f
Q^2_f \left\{ (z^2 + (1-z)^2) \epsilon^2 K_1 (\epsilon r_\bot)^2 + 
m^2_f K_0(\epsilon r_\bot )^2 \right\}
\label{(29)}
\ee
and 
\be
\left| \psi_{L} (r_\bot , z (1-z) , Q^2 ) \right|^2 = \frac{6 \alpha}{(2\pi)^2} 
\sum_f Q^2_f 4 Q^2 z^2 (1-z)^2 K_0 (\epsilon r_\bot)^2
\label{(30)}
\ee
where
\be
\epsilon^2 = z (1 - z) Q^2 + m^2_f  .
\label{(31)}
\ee
In (\ref{(29)}) to (\ref{(31)}), $Q_f$ denotes the quark charge in units of 
$e , \alpha = e^2/4\pi$ the electromagnetic fine-structure constant, and 
$m_f$ denotes the quark mass. The functions $K_0(\epsilon r_\bot)$ and 
$K_1 (\epsilon r_\bot)$ are modified Bessel functions. 

It is important to emphasize that the justification for the use of 
the representations 
(\ref{(27)}) and (\ref{(28)}) as the $x\rightarrow 0$ limit of the 
two-gluon-exchange mechanism rests on applying them in conjunction with the 
representation for the color-dipole cross section given by \cite{b,e} 
\begin{eqnarray}
\sigma_{(q \bar q)p} (r_\bot , W^2) & = & 
\int d^2 l_\bot \tilde\sigma_{(q \bar q)p} \left( \vec l^{~2}_\bot , 
W^2 \right) \left( 1 - e^{-i \vec l_\bot \cdot \vec r_\bot} \right) \nonumber \\
& = & \sigma^{(\infty)}\cdot \left\{ \matrix{
\frac{1}{4} r^2_\bot \langle \vec l^{~2}_\bot \rangle_{W^2} , & {\rm for} ~
\vec r^{~2}\langle \vec l^{~2}_\bot \rangle_{W^2} \rightarrow 0 , \cr 
1  & ~{\rm for} ~ \vec r^{~2} \rightarrow \infty , \cr }   \right.
\label{(32)}
\end{eqnarray} 
where by definition 
\be
\langle \vec l^{~2}_\bot \rangle_{W^2} = \frac{\int d \vec l^{~2}_\bot
\vec l^{~2}_\bot\tilde\sigma\left(\vec l^{~2}_\bot,W^2 \right)}
{\int d \vec l^{~2}_\bot \tilde\sigma\left(\vec l^{~2}_\bot,W^2 \right)}
\label{(33)}
\ee
and
\be
\sigma^{(\infty)} = \pi \int d \vec l^{~2}_\bot \tilde\sigma \left( 
\vec l^{~2}_\bot , W^2 \right) . 
\label{(34)}
\ee
The limit of $\vec r^{~2}_\bot \rightarrow 0$ corresponds to a vanishing 
interaction strength, due to color neutrality of the $q \bar q$ color-dipole
state. The finite limit, $\sigma^{(\infty)}$, avoids an infinite color-dipole 
cross section in the limit of infinite quark-antiquark separation
\footnote{It is worth noting that the energy dependence of 
the color-dipole cross section in the GVD/CDP\cite{e} enters 
exclusively via a (soft) 
increase of the gluon-transverse-momentum transfer, $\vec l_\bot$,
with energy, i.e., via $\tilde\sigma_{(q\bar q)p}(\vec l^2_\bot)
=\tilde\sigma_{(q\bar q)p}(\vec l^2(W^2))$.}, and it 
guarantees hadronic unitarity, provided $\sigma^{(\infty)}$ is well-behaved 
for $W\rightarrow \infty$.

For the ensuing examination of the spin properties of the $q \bar q$ states, it 
proves useful to introduce the variable
\be
\vec r^{~\prime}_\bot = \sqrt{z (1 - z)} \vec r , 
\label{(35)}
\ee
and to first of all consider the integral over 
$d^2 r^\prime_\bot$ of the $q \bar q$-vacuum-polarization 
loop by itself. \footnote{The relevance of hadronic
vacuum polarization for photo- and electroproduction 
(from nuclei) at low $x$ 
during the pre-QCD era was stressed in particular by V. Gribov \cite{x}} 
We restrict ourselves to a 
vanishing quark mass, $m_f = 0$. \footnote{In applying the approach based on 
(\ref{(27)}) to the description of the experimental data, it proved useful 
\cite{e} to 
return to momentum space and to introduce an effective constituent-quark mass 
that coincides with the value suggested by quark-hadron duality
\cite{Sakurai} in $e^+e^-$-annihilation. 
This mass then effectively provides a lower limit in the 
integration over the mass spectrum of the $q \bar q$ vector states the photon 
virtually dissociates into.}
In terms of $\vec r^{~\prime}_\bot$ from (\ref{(35)}), we have 
\begin{eqnarray}
\int d^2 r_\bot \left| \psi_T (r_\bot ,z (1-z) , Q^2) \right|^2 & = & 
\int \frac{d^2 r^{~\prime}_\bot}{z (1-z)} \left| \psi_T \left( \frac{r^{~\prime}_
\bot}{\sqrt{z (1-z)}} , z(1-z) , Q^2 \right) \right|^2 
\label{(36)}   \\
& = & \frac{6\alpha}{(2\pi)^2} Q^2 \sum_f Q^2_f (z^2 +(1-z)^2)\int d^2 
r^{~\prime}_\bot K_1 \left(\sqrt{Q^2} r^{~\prime}_\bot \right)^2  , 
\nonumber
\end{eqnarray} 
and
\begin{eqnarray}
\int d^2 r_\bot \left| \psi_L (r_\bot , z(1-z) , Q^2) \right|^2 & = &     
\int \frac{d^2 r^{~\prime}_\bot}{z (1-z)} \left| \psi_L \left( \frac{r^{~\prime}_
\bot}{\sqrt{z (1-z)}} , z(1-z) , Q^2 \right) \right|^2 
\label{(37)}    \\
& = & \frac{6\alpha}{(2\pi)^2} Q^2 \sum_f Q^2_f z (1-z) \int d^2 r^{~\prime}_\bot
K_0 \left( \sqrt{Q^2} r^{~\prime}_\bot \right)^2 .~~~~~~~~~ \nonumber 
\end{eqnarray}
The $z$-dependence in (\ref{(36)}) and (\ref{(37)}) that originates from the 
$\gamma^* q \bar q$ coupling is immediately seen to be characteristic of the 
spin $J=1$ nature of the photon. It coincides with the angular 
dependence in the $q\bar q$ rest frame well-known
from e.g. $e^+ e^-$ annihilation into $q \bar q$. Indeed, upon identifying 
\begin{eqnarray}
       \sin^2 \theta & = & 4z (1-z) , \nonumber \\
       \cos \theta &=& 1 - 2 z , 
\label{(8)}
\end{eqnarray}
we may represent the integrations over $dz$ in (\ref{(27)}) in terms 
of $\cos\theta$ and the 
rotation function $d^1_{\lambda ,\mu}(z)$,
\begin{eqnarray}
d^1_{1,1} (z) & = & \frac{1}{2} (1 + \cos\theta) = 1 - z , \nonumber \\
d^1_{1,-1} (z) &=& \frac{1}{2} (1-\cos\theta) = z , \label{(101)}\\
d^1_{1,0}(z) &=& \frac{1}{\sqrt 2}\sin\theta = \sqrt2 \sqrt{z(1-z)} . 
\nonumber 
\end{eqnarray}
The integrations over $dz$ in (\ref{(27)}) and (\ref{(28)}), 
as far as the photon wave functions 
are concerned, according to (\ref{(36)}) and (\ref{(37)}) can then be written as 
\begin{eqnarray}
dz \cdot  (z^2 +(1-z)^2) & = & 
= - \frac{1}{2} d \cos \theta \cdot \frac{1}{2} (1 + 
\cos^2 \theta) \nonumber \\
 & = & dz \cdot \left[ | d^1_{1,1} (z)|^2 + | d^1_{1,-1} (z) |^2 \right]
\label{(102)}  
\end{eqnarray}
and
\begin{eqnarray}
dz \cdot z (1-z) & = & - \frac{1}{2} d \cos \theta \cdot \frac{1}{4} \sin^2
\theta \nonumber \\
& = & dz \cdot \frac{1}{2} | d^1_{1,0} (z) |^2 . 
\label{(103)} 
\end{eqnarray}
In (\ref{(102)}), we recognize the $(1+\cos^2 \theta)$ distribution from 
$e^+ e^-$ annihilation into $q \bar q$. 

Returning to the virtual photoabsorption cross section (\ref{(27)}) and 
introducing 
$\vec r^{~\prime}_\bot$, we have 
\begin{eqnarray}
 & &\sigma_{\gamma^*_{T,L} p} (W^2 , Q^2) \label{(38)}\\
& & = \int dz \int \frac{d^2 r^{~\prime}
_\bot}{z(1-z)} 
\left| \psi_{T,L} \left( \frac{r^{~\prime}_
\bot}{\sqrt{z (1-z)}} , z(1-z) , Q^2 \right) \right|^2 
\sigma_{(q \bar q)p} \left(\frac{r^{~\prime}_\bot}{\sqrt{z (1-z)}}, 
 W^2\right) ,
\nonumber
\end{eqnarray}
where according to (\ref{(32)}) 
\begin{eqnarray}
& & \sigma_{(q \bar q)p} (\frac{r^{~\prime}_\bot}{\sqrt{z (1-z)}}, W^2) 
\nonumber \\
& & = 
\int d^2 l^\prime_\bot z (1-z) 
\tilde\sigma_{(q\bar q)p} \left( \vec l^{~\prime 2}_\bot z (1-z) , W^2 \right) 
\left( 1 - e^{-i \vec l^{~\prime}_\bot \vec r^{~\prime}_\bot} \right) . 
\label{(39)}
\end{eqnarray}
In (\ref{(39)}), we use 
\be
\vec l^{~\prime}_\bot = \frac{\vec l_\bot}{\sqrt{z(1-z)}} . 
\label{(104)}
\ee
Reading (\ref{(38)}) in conjunction with (\ref{(36)}) and (\ref{(37)})
as well as (\ref{(39)}), we note the appearance of products of 
$d^1_{\lambda ,\mu} (z)$ functions in (\ref{(38)}). 
In order to take advantage of their 
orthogonality properties, we expand the $(q \bar q)p\rightarrow (q \bar q)p$
amplitude in (\ref{(39)}) \footnote{Actually, we expand the amplitude for
$(q \bar q)^{J=1} p \rightarrow (q \bar q)^J p$ in terms of the spin $J$ of 
the outgoing $(q \bar q)$ state. This is seen in (\ref{(11)}) and (\ref{(12)}), 
where $z(1-z) \tilde\sigma (\vec l^{~\prime 2}_\bot z(1-z), W^2)$ is 
multiplied by an appropriate projection factor stemming from the 
substitution of (\ref{(39)}) into (\ref{(38)}).}
in terms of states of different spin $J$ of the $q\bar q$ system, i.e. in terms
of $\bar\sigma_{(q \bar q)^J_{T,L}p} (\vec l^{~\prime 2}_\bot , W^2)$.
Accordingly, 
\begin{eqnarray}
  & & (1-z) \left[ z(1-z) \tilde\sigma_{(q\bar q)p}
(\vec l^{~\prime 2}_\bot z 
(1-z),W^2) \right]              \nonumber \\
  &=& d^1_{11} (z) \bar\sigma_{(q \bar q)^{J=1}_T p}
                   (\vec l^{~\prime 2}_\bot , W^2) 
     +d^2_{11} (z) \bar\sigma_{(q \bar q)^{J=2}_T p}
                   (\vec l^{~\prime 2}_\bot , W^2) +\cdots , \nonumber \\
  & & z \left[ z (1-z)\tilde\sigma_{(q\bar q)p} (\vec l^{~\prime 2}_\bot z 
(1-z),W^2)  \right]            \label{(11)}\\
  &=& d^1_{1-1} (z) \bar\sigma_{(q \bar q)^{J=1}_T p}
                   (\vec l^{~\prime 2}_\bot , W^2) 
     -d^2_{1-1} (z) \bar\sigma_{(q \bar q)^{J=2}_T p}
 (\vec l^{~\prime 2}_\bot , W^2) +\cdots \nonumber 
\end{eqnarray}
and
\begin{eqnarray}
& & \sqrt 2 \sqrt{z(1-z)} z(1-z)  \tilde\sigma_{(q \bar q) p}
(\vec l^{~\prime 2}_\bot z (1-z) , 
W^2) \nonumber\\
&= & d^1_{10} (z) \bar\sigma_{(q \bar q)^{J=1}_L p} (\vec l^{~\prime 2}_\bot , 
W^2) + d^2_{10} (z)
\bar\sigma_{(q \bar q)^{J=2}_L p} (\vec l^{~\prime 2}_\bot , W^2) + \cdots .
\label{(12)}
\end{eqnarray} 
Inserting (\ref{(11)}) and (\ref{(12)}), respectively, into (\ref{(39)})
and subsequently (\ref{(39)}) into (\ref{(38)}), we note that indeed only 
contributions due to elastic interactions $(q \bar q)^{J=1}_{T,L} p \rightarrow 
(q \bar q)^{J=1}_{T,L}p$, with the 
proton remain, and (\ref{(38)}) becomes 
\be
       \sigma_{\gamma^*_{T,L}p} (W^2 , Q^2) 
     = \int dz \int \frac{d^2 r^\prime_\bot}{z(1-z)} | \psi_{T,L} 
       ( \frac{r^{~\prime}_\bot}{\sqrt{z (1-z)}}, z(1-z), Q^2) |^2 
       \sigma_{(q \bar q)^{J=1}_{T,L}p}(\vec r^{~\prime}_\bot , W^2) .
\label{(41)}
\ee
According to (\ref{(39)}), the cross sections 
$\sigma_{(q \bar q)^{J=1}_{T,L} p}(\vec r_\bot^{~\prime},W^2)$ in
(\ref{(41)}), 
corresponding to the (imaginary parts of the) elastic 
forward-scattering amplitudes  
$(q \bar q)^{J=1}_{T,L} p \rightarrow 
(q \bar q)^{J=1}_{T,L}p$, are given by
\be
\sigma_{(q \bar q)^{J=1}_{T,L}p } (\vec r^{~\prime}_\bot , W^2) = 
\int d^2 l^\prime
_\bot \bar\sigma_{(q \bar q)^{J=1}_{T,L} p} 
( \vec l^{~\prime 2}_\bot , W^2) \left(1 - e^{-i 
\vec l^{~\prime}_\bot \cdot \vec r^{~\prime}_\bot} \right) ,
\label{(40)}
\ee
where $\bar\sigma_{q \bar q)^{J=1}_{T,L}p}(\vec l^{~\prime 2}_\bot , W^2)$ 
according to (\ref{(11)}) and (\ref{(12)}) is explicitly given by 
\begin{eqnarray}
   \bar\sigma_{(q \bar q)^{J=1}_T p} (\vec l^{~\prime 2}_\bot , W^2) 
   &=& 3\int dz  z(1-z)^3 \tilde\sigma_{(q \bar q) p} 
       (\vec l^{~\prime 2}_\bot z (1-z), W^2) \nonumber \\
   &=& 3\int dz  z^3(1-z) \tilde\sigma_{(q \bar q) p} 
       (\vec l^{~\prime 2}_\bot z (1-z), W^2)  \label{(201)}
\end{eqnarray}
and 
\be
     \bar\sigma_{(q \bar q)^{J=1}_L p} (\vec l^{~\prime 2}_\bot , W^2) 
    = 6 \int dz (z(1-z))^2 \tilde\sigma_{(q \bar q)p} 
      (\vec l^{~\prime 2}_\bot z (1-z), W^2) .
\label{(202)}
\ee
We stress that (\ref{(41)}) with (\ref{(40)}) is no less 
general than (\ref{(27)}) with (\ref{(32)}). 
The difference between these two representations is only 
due to the exploitation of the decomposition of the 
$(q \bar q)^{J=1}p \rightarrow (q \bar q)^J p$ amplitude 
into contributions from 
different spin $J$ of the $q \bar q$ system. 

The representation (\ref{(27)}) with (\ref{(32)}) for the total cross section
$\sigma_{\gamma^*_{T,L}p}(W^2,Q^2)$ contains redundant and irrelevant 
contributions on the right-hand side due to $(q \bar q)^{J=1} p
\rightarrow (q \bar q)^{J\not= 1} p$ scattering amplitudes. These contributions
vanish upon integration, and they are eliminated right from the beginning in the 
representation (\ref{(41)}) with  (\ref{(40)}).

Fits to the experimental data for $\sigma_{\gamma^*_{T,L}p}(W^2 , Q^2)$ are 
frequently based \cite{k,l,m} on  (\ref{(27)}).
The representation 
\be
\sigma_{\gamma^*_{T,L} p} (W^2 , Q^2 ) = \int dz \int d^2 r_\bot \left| 
\psi_{T,L} (r_\bot , z (1-z) , Q^2 ) \right|^2 
\sigma_{(q \bar q )p} (\vec r_\bot , W^2 ) 
\label{(105)}
\ee
is used for the fit \cite{k,l,m} upon adopting an ansatz for the 
color-dipole cross section, 
$\sigma_{(q \bar q)p}$, as a function of $r_\bot$ and either $W^2$ \cite{k}, 
or alternatively,  $x$ \cite{l}. 
Different functional forms for $\sigma_{(q \bar q)p}$ will in general yield
equally good fits for their specific sets of fit parameters, provided the 
different functional forms agree in their $(q \bar q)^{J=1} p
\rightarrow (q \bar q)^{J = 1} p$ content. 
It is of no surprise that widely differing fit  
results \cite{n} for color-dipole cross sections were extracted by 
different authors; dipole-cross sections 
that implicitly contain approximately identical 
$(q \bar q)^{J=1}$ but widely differing $(q \bar q)^{J\not= 1}$ final-state
contributions obviously yield identical representations of the total 
cross section, $\sigma_{\gamma^* p}(W^2, Q^2)$. 

Such ambiguities are avoided in the representation (\ref{(41)}) 
with (\ref{(40)}). 
For the sake of clarity, it may be appropriate to equivalently express 
 (\ref{(41)}) in terms of the original variable $r_\bot$, 
\begin{eqnarray}
& & \sigma_{\gamma^*_{T,L}p} (W^2 , Q^2) = \label{(42)} \\
& & = \int dz \int d^2 r_\bot 
| \psi_{T,L} (r_\bot , z(1-z), Q^2) |^2 
\sigma_{(q \bar q)^{J=1}_{T,L}p}
(\vec r_\bot \sqrt{z(1-z)}, W^2) 
\nonumber
\end{eqnarray} 
where according to (\ref{(40)}), 
\be
\sigma_{(q \bar q)^{J=1}_{T,L}} (\vec r_\bot \sqrt{z(1-z)}, W^2) = \int d^2 l^
\prime_\bot \bar\sigma_{(q \bar q)^{J=1}_{T,L}p} (\vec l^{~\prime 2}_\bot , 
W^2)(1 - e^{-i\vec l^{~\prime}_\bot \cdot \vec r_\bot \sqrt{z(1-z)}}) .
\label{(108)}
\ee
In distinction from (\ref{(105)}), in  (\ref{(42)}) the variable $r^\prime_\bot
= r_\bot \cdot \sqrt{z(1-z)}$ appears as argument of the dipole cross section, and
this is sufficient to assure that only $(q \bar q)^{J=1} p \rightarrow 
(q \bar q)^{J=1}p$ is included. 

A few additional comments on the form  (\ref{(41)}) for 
$\sigma_{\gamma^*_{T,L}p}$ may be appropriate. 
The dipole cross section in  (\ref{(41)}) depends only on 
$\vec r^{~\prime}$ and $W^2$, while the dependence on $Q^2$ and the 
$q \bar q$ configuration variable $z$ is transferred 
to the photon wave function thus describing the $q \bar q$ 
fluctuation of the photon with the appropriate $z$ dependence 
as in $e^+e^-$ annihilation; compare  the  second
equality in (\ref{(36)}) and (\ref{(37)}). 
Once $\sigma_{(q \bar q)^{J=1}}
(r^\prime_\bot , W^2)$ is determined from a fit to the experimental data for  
$\sigma_{\gamma^*_{T,L}p} (W^2 , Q^2)$, it may be inserted into the cross 
section for diffractive production (\ref{(28)}) via 
\be
\sigma_{(q \bar q)p} (\vec r_\bot , W^2) \rightarrow 
\sigma_{(q \bar q)^{J=1}_{T,L}p}
(\vec r_\bot \sqrt{z(1-z)} , W^2)
\label{(43)}
\ee
to obtain a unique prediction for ``elastic'' diffractive forward production
\be
      \gamma^*_{T,L} p \rightarrow (q \bar q)^{J=1}_{T,L} p
\label{(106)}
\ee
of $q \bar q$ vector states. It is diffractive production of vector states 
that is uniquely connected with the total cross section 
$\sigma_{\gamma^*p} (W^2 , Q^2)$ at low $x$. 
Besides the $J=1$ diffractive continuum, via quark-hadron duality
\cite{Sakurai, martin}, it is 
vector meson production in particular that is uniquely related 
to the total cross section $\sigma_{\gamma^*_{T,L} p}(W^2 , Q^2)$. 

Diffractive production in general contains a large part of ``inelastic'' 
diffraction,\footnote{Compare\cite{bialas} for a treatment of the
total cross section and of diffraction based on an elastic and
inelastic component.} 
\be
    \gamma^*_{T,L} p \rightarrow (q \bar q)^{J\not= 1} p,
\label{(107)}
\ee
that remains unconstrained by fits of the dipole cross section to 
$\sigma_{\gamma^*_{T,L}p}(W^2,Q^2)$. A description of the sum of elastic 
and inelastic diffractive 
production according to (\ref{(28)}) must contain an additive component
in the dipole cross section that is projected to zero and thus remains inert 
when included in (\ref{(27)}).  

In short, if (\ref{(41)}) with (\ref{(40)}) is used to describe the total 
cross section, a successful description will imply a prediction for 
elastic diffraction, (\ref{(106)}). 
If (\ref{(27)}) with (\ref{(32)}) is employed to describe the total cross 
section by fitting an ansatz for the dipole cross section, it is by no means
guaranteed that this fit, when substituted into (\ref{(28)}), will be 
relevant for a representation of elastic as well as inelastic diffractive 
production. After all, as repeatedly stressed, inelastic diffraction remains 
unconstrained by $\sigma_{\gamma^*_{T,L} p} (W^2,Q^2)$. 
The use of the form (\ref{(27)}) for the total cross section necessitates 
a simultaneous fit to both, the total cross section (\ref{(27)}) and 
(the sum of elastic and inelastic) 
diffractive production according to (\ref{(28)}). 
In principle\footnote{Since the data on diffraction do not 
reach the accuracy of the data for $\sigma_{\gamma^* p}$, 
it seems preferable to fit $\sigma_{\gamma^* p}$ according 
to (\ref{(41)}) with (\ref{(40)}) and compare with elastic 
diffraction (into continuum states as well as vector-meson production) 
and subsequently fit inelastic diffraction
by introducing an (orthogonal) additive contribution.}, 
such a fit will provide 
a unique color-dipole cross section that in addition to $\sigma_{\gamma^*_{
T,L}p}(W^2 , Q^2)$ describes both elastic and inelastic diffraction. 
If the fit to diffraction is achieved by an additive contribution 
\cite{k,l,m} in the dipole
cross section relative to the one successfully used in the fit to 
$\sigma_{\gamma^*_{T,L}p}(W^2 , Q^2)$, it ought to be verified that the 
added term only contributes to inelastic diffraction while leaving the 
(previous) fit to $\sigma_{\gamma^*_{T,L}p}(W^2 , Q^2)$ unchanged.

\section{Momentum space representation and sum rule.}

\setcounter{equation}{0}

\renewcommand{\theequation}{%
\mbox{\arabic{section}.\arabic{equation}}}

The connection between the total cross section and elastic diffraction 
becomes explicit in terms of a sum rule \cite{d} that relates the total cross
section to elastic diffractive forward production. 

In momentum space, the $x\rightarrow 0$ representation of the total 
cross section (\ref{(27)}) becomes \cite{d} 
\begin{eqnarray}
\sigma_{\gamma^*_{T,L} p} (W^2 , Q^2) = & & 
\frac{3}{16\pi^3 \cdot 2} \int dz  \int d^2 k_\bot
\int d^2 l_\bot \tilde\sigma_{(q \bar q)p}
(\vec l^{~2}_\bot , W^2)  \cdot \nonumber \\
& & \cdot | {\cal M}_{T,L} (z, \vec k_\bot , Q^2) - {\cal M}_{T,L} (z, \vec k_
\bot + \vec l_\bot , Q^2) |^2 . 
\label{(301)}
\end{eqnarray}
The representation (\ref{(301)}) is related to the representation in transverse 
position space (\ref{(27)}) by the substitution of the photon wave 
function in momentum space that is given by 
\begin{eqnarray}
\sum_{\lambda,\lambda^\prime}\Big| \psi_{T,L}^{(\lambda, \lambda^\prime)} 
(\vec r_\bot , z; Q^2) \Big|^2 = 
& & 3 \cdot \frac{4\pi}{(16\pi^3)^2} \int d^2 k^{~\prime}_\bot 
\int d^2
k_\bot {\cal M}^*_{T,L} (\vec k^{~\prime}_\bot , z , Q^2) 
\nonumber  \\        
& & {\cal M}_{T,L} (\vec k_\bot , z ,Q^2) 
\exp (i \vec k^{~\prime}_\bot - \vec k_\bot ) \vec r_\bot , 
 \label{(302)} 
\end{eqnarray}
with 
\be
{\cal M}^*_T (\vec k^{~\prime}_\bot , z, Q^2) \cdot {\cal M}_T (\vec k_\bot, 
z, Q^2) = \frac{8\pi \alpha (\vec k^{~\prime}_\bot \cdot \vec k_\bot )
\sum_f Q^2_f (z^2 + (1 - z)^2)}{(z (1-z)Q^2 + \vec k^{~
\prime 2}_\bot) (z (1-z) Q^2 + \vec k^{~2}_\bot)}
\label{(303)}
\ee
and
\be
{\cal M}^*_L (\vec k^{~\prime}_\bot , z, Q^2) \cdot {\cal M}_L (\vec k_\bot, 
z, Q^2) = \frac{32\pi \alpha Q^2 \sum_f Q^2_f z^2  (1 - z)^2}
{(z (1-z)Q^2 + \vec k^{~
\prime 2}_\bot )(z (1-z) Q^2 + \vec k^{~2}_\bot)}.
\label{(304)}
\ee
In (\ref{(301)}), $\vec k_\bot$ denotes the transverse momentum of the quark. 
It is related to the mass, $M$, of the $q \bar q$ state by 
\be
M^2 = \frac{\vec k^{~2}_\bot + m^2}{z(1-z)} ,
\label{(305)}
\ee
where the quark mass $m$ will be put to $m=0$ for simplicity. After one 
angular integration, and upon introducing $M^2$ as integration variable, 
via 
\be
d^2 k_\bot = \frac{1}{2} z(1-z) dM^2 d\varphi , 
\label{(315)}
\ee 
(\ref{(301)}) becomes\footnote{Note that we introduce \cite{e} 
the threshold mass 
$m_0$, as announced in the footnote in connection with (\ref{(36)}).
  We ignore the additive "correction  terms" given in \cite{e, d}
that assure an identical threshold mass, $m_0$ for the incoming and
outgoing $q\bar q$  pair in the forward Compton amplitude.}
\begin{eqnarray}
& & \sigma_{\gamma^*_T p} (W^2 , Q^2) = \frac{c_0^2}{64\pi}    \int_{m^2_0} d M^2 
   \frac{M^2}{(Q^2 + M^2)^2} \int dz (z^2+(1-z)^2)    \nonumber   \\
   & & \cdot \int d^2 l^\prime_\bot z(1-z) \tilde\sigma_{(q \bar q)p} 
   (\vec l^{~\prime 2}_\bot z (1-z), 
   W^2) F_T (Q^2 , M^2 , \vec l^{~\prime 2}_\bot ), 
\label{(306)} 
\end{eqnarray}
and
\begin{eqnarray}
& & \sigma_{\gamma^*_L p} (W^2 , Q^2) = \frac{c_0^2}{16\pi} \int_{m^2_0} d M^2 
\frac{Q^2}{(Q^2 + M^2)^2} \int dz z(1-z)    \nonumber   \\
& & \cdot \int d^2 l^\prime_\bot z(1-z) \tilde\sigma_{(q \bar q)p}
(\vec l^{~\prime 2}_\bot z (1-z), 
W^2)\cdot F_L (Q^2 , M^2 , \vec l^{~\prime 2}_\bot ). 
\label{(307)} 
\end{eqnarray}
Here,
\be
c_0 = 2 \sqrt{\frac{2N_c}{\pi}} \sqrt{\frac{4\pi\alpha R_{e^+ e^-}}{3}} .
\label{(308)}
\ee
The number of quark colors is $N_c = 3$, and $R_{e^+ e^-}$ denotes the 
cross section for $e^+ e^-$ annihilation, $e^+ e^- \rightarrow q \bar q 
\rightarrow$ hadrons, in units of the cross section for 
$e^+ e^- \rightarrow \mu^+ \mu^-$, 
\be
R_{e^+ e^-} = 3 \sum_f Q^2_f .
\label{(309)}
\ee 
The sum runs over the quark charges, $Q_f$, in units of $e$.
The functions $F_T (Q^2 , M^2 , \vec l^{~\prime 2}_\bot )$ and 
$F_L (Q^2 , M^2 , \vec l^{~\prime 2}_\bot )$
in (\ref{(306)}) and (\ref{(307)}) 
are given by 
\be
F_T (Q^2, M^2, \vec l^{~\prime 2}_\bot) = 1 - \frac{Q^2 + M^2}{2M^2} \left(
1 + \frac{M^2 - Q^2 - \vec l^{~\prime 2}_\bot}{\sqrt{(Q^2 + M^2 + 
\vec l^{~\prime 2}_\bot)^2 - 4M^2 \vec l^{~\prime 2}_\bot}} \right)
\label{(310)}
\ee 
and 
\be
F_L (Q^2, M^2, \vec l^{~\prime 2}_\bot) = 1 - \frac{Q^2 + M^2}{\sqrt{(M^2 + Q^2 +
\vec l^{~\prime 2}_\bot )^2 - 4 M^2 \vec l^{~\prime 2}_\bot }} .
\label{(311)}
\ee 
The $z$-dependence in (\ref{(306)}) and (\ref{(307)})
is identical to the one encountered 
in transverse-position-space. 
As in the transverse-position-space treatment, we now use (\ref{(11)}) and 
(\ref{(12)}) and express the transverse and the longitudinal photoabsorption
cross section in (\ref{(306)}) and (\ref{(307)}), respectively, in terms of 
$\bar\sigma_{(q \bar q)^{J=1}_{T,L}p}(\vec l^{~\prime 2} , W^2)$ from 
(\ref{(201)}) and (\ref{(202)}). We obtain
\begin{eqnarray}
& & \sigma_{\gamma^*_T p} (W^2 , Q^2) = \frac{c_0^2}{64\pi} \int_{m^2_0}
dM^2 \frac{M^2}{(Q^2 + M^2)^2} \cdot \nonumber \\
& & \cdot \int dz (z^2 +(1-z)^2) \int d^2 l^\prime_\bot \bar\sigma_{(q \bar q)
^{J=1}_T p} 
(\vec l^{~\prime 2}_\bot , W^2) F_T (Q^2 , M^2 , \vec l^{~\prime 2}_\bot) 
\label{(312)}
\end{eqnarray}
and
\begin{eqnarray}
& & \sigma_{\gamma^*_L p} (W^2 , Q^2) = \frac{c_0^2}{16\pi} \int_{m^2_0}
dM^2 \frac{Q^2}{(Q^2 + M^2)^2} \cdot \nonumber \\
& & \cdot \int dz z (1-z) \int d^2 l^\prime_\bot \bar\sigma_{(q \bar q)^
{J=1}_L p} 
(\vec l^{~\prime 2}_\bot , W^2) F_L (Q^2 , M^2 , \vec l^{~\prime 2}_\bot)  .
\label{(313)}
\end{eqnarray}
The integrations over $z$ in (\ref{(312)}) and (\ref{(313)}) can be carried out 
immediately to yield the factors $2/3$ and $1/6$, respectively. 

The results (\ref{(312)}) and (\ref{(313)}) may be conveniently 
summarized by the substitution rule 
\be
z(1-z) \tilde\sigma_{(q \bar q)p} (\vec l^{~\prime 2}_\bot z(1-z), W^2)
\rightarrow \bar\sigma_{(q \bar q)^{J=1}_{T,L} p} (\vec l^{~\prime 2}_\bot , W^2)
\label{(314)}
\ee 
to be applied in (\ref{(306)}) and (\ref{(307)}). We note that 
(\ref{(312)}) and (\ref{(313)}) are the analogue of the 
transverse-position-space result (\ref{(41)}). 
According to (\ref{(312)}) and (\ref{(313)}), the total cross section, 
$\sigma_{\gamma^*_{T,L}p} (W^2 , Q^2)$, or equivalently the virtual forward 
Compton scattering amplitude, for $x \ll 1$, is determined by 
$(q \bar q)^{J=1}_{T,L} p \rightarrow (q \bar q)^{J=1}_{T,L}p$
forward scattering. 

We turn to diffractive production and the derivation of the sum rule. We 
substitute the momentum-space expressions of the photon-wave functions 
(\ref{(302)}) to (\ref{(304)}) into (\ref{(28)}), as well as the dipole 
cross section (\ref{(32)}). Introducing $M^2$ from (\ref{(305)}) 
the cross section for diffractive production, 
$\gamma^*_{T,L}p \rightarrow X_{T,L} p$, reads 
\begin{eqnarray}
      && \left.\frac{d\sigma_{\gamma^*_{T,L}p \rightarrow X_{T,L}p}}
        {dt dz dM^2} \right|_{t=0}= 
        \frac{3}{2} \frac{1}{(16\pi^2)^2}   \label{(316)} \\
      &&\cdot z (1-z) \int d \varphi 
        \left[ \int d^2 l_\bot 
        \tilde\sigma_{(q \bar q)p} ( \vec l^{~2}_\bot, W^2) 
        ({\cal M}_{T,L} (\vec k_\bot , z, Q^2) - 
         {\cal M}_{T,L} (\vec k_\bot + \vec l_\bot , z, Q^2)) 
         \right]^2 . \nonumber
\end{eqnarray}
Upon angular integration, (\ref{(316)}) for transverse and 
longitudinal photons respectively, leads to  
\begin{eqnarray}
& & \sqrt{\left.\frac{d\sigma_{\gamma^*_T p\rightarrow Xp}}
{dz dt dM^2}\right|_{t=0}} =  \frac{c_0}{32\pi}
\sqrt{z^2+(1-z)^2} \frac{M}{Q^2 + M^2} \cdot \label{(317)} \\
& & \cdot \int d^2 l^\prime_\bot z(1-z) \tilde\sigma
(\vec l^{~\prime 2}_\bot z (1-z), W^2) F_T(Q^2 , M^2 , 
\vec l^{~\prime 2}_\bot) , \nonumber
\end{eqnarray}
and
\begin{eqnarray}
& & \sqrt{\left.\frac{d\sigma_{\gamma^*_L p\rightarrow Xp}}
{dz dt dM^2}\right|_{t=0}} =  \frac{c_0}{16\pi}
\sqrt{z(1-z)} \frac{\sqrt{Q^2}}{Q^2 + M^2} \cdot \label{(318)} \\
& & \cdot \int d^2 l^\prime_\bot z(1-z) \tilde\sigma
(\vec l^{~\prime 2}_\bot z (1-z), W^2) F_L (Q^2 , M^2 , 
\vec l^{~\prime 2}_\bot) . \nonumber
\end{eqnarray}
Finally, we compare (\ref{(317)}) and (\ref{(318)}) with 
(\ref{(306)}) and (\ref{(307)}) to obtain the sum rules 
\begin{eqnarray}
&&\sigma_{\gamma^*_T p} (W^2 , Q^2) ={{c_0}\over 2} \int_{m^2_0} dM^2 
\frac{M}{Q^2 + M^2} 
\cdot \nonumber \\
&&\cdot \int dz \sqrt{z^2+(1-z)^2} \sqrt{\left.\frac{d\sigma_{\gamma^*_T p
\rightarrow X_{T,L} p}}{dz dt dM^2}\right|_{t=0}} 
\label{(319)} 
\end{eqnarray}
and 
\begin{eqnarray}
&&\sigma_{\gamma^*_L p} (W^2 , Q^2) = c_0 \int_{m^2_0} dM^2 
\frac{\sqrt{Q^2}}{Q^2 + M^2} \cdot \nonumber \\
&&\cdot \int dz \sqrt{z(1-z)} \sqrt{\left.\frac{d\sigma_{\gamma^*_L p
\rightarrow X_{T,L}p}}{dz dt dM^2}\right|_{t=0}}   .
\label{(320)} 
\end{eqnarray}
Note that the diffractive production cross sections in general involve final 
states $X$ that contain $(q\bar q)^J$ states of arbitrary spin $J$. The 
corresponding distribution in $z$ should appear as angular 
distribution of a quark and an antiquark jet in the rest frame of the 
$(q \bar q)^J$ system. The $z$-dependent projections in 
(\ref{(319)}) and (\ref{(320)}), according to (\ref{(102)}) and (\ref{(103)}), 
project on $J=1$ final states, $(q \bar q)^{J=1}$. 

We may rewrite (\ref{(319)}) and (\ref{(320)}) as 
\footnote{The sum rules (\ref{(321)}) and (\ref{(322)}) are identical
to the ones given in \cite{d}.  As repeatedly stressed,
in \cite{d} they were obtained upon introducing a
specific ansatz  for the color-dipole cross section \cite{e},
while the present derivation is entirely based on the 
two-gluon-exchange generic structure.}
\be
\sigma_{\gamma^*_T p} (W^2 , Q^2) = \sqrt{16\pi} 
\sqrt{\frac{\alpha R_{e^+ e^-}}{3\pi}} \int_{m^2_0} dM^2 \frac{M}{Q^2 + M^2} 
\sqrt{\left.\frac{d\sigma_{\gamma^*_T p
\rightarrow X^{J=1}_T p}}{dt dM^2}\right|_{t=0}} 
\label{(321)} 
\ee
and
\be
\sigma_{\gamma^*_L p} (W^2 , Q^2) = \sqrt{16\pi} 
\sqrt{\frac{\alpha R_{e^+ e^-}}{3\pi}} \int_{m^2_0} dM^2 \frac{\sqrt{Q^2}}
{Q^2 + M^2} 
\sqrt{\left.\frac{d\sigma_{\gamma^*_L p
\rightarrow X^{J=1}_L p}}{dt dM^2}\right|_{t=0}} .
\label{(322)} 
\ee
The sum rules (\ref{(321)}) and (\ref{(322)}) express the total cross section
in terms of elastic diffraction, $\gamma^*_{T,L}p \rightarrow X^{J=1}_{T,L} p$, 
where $X^{J=1}_{T,L}$ stands for $X^{J=1}_{T,L} \equiv (q \bar q)^{J=1}_{T,L}$.
Explicitly, the diffractive-production cross sections in (\ref{(321)}) and 
(\ref{(322)}) are given by 
\be
\left. \frac{d\sigma_{\gamma^*_T p \rightarrow X^{J=1}_T p}}{dt dM^2} 
\right|_{t=0} = \Bigl({{c_0}\over{32\pi}}\Bigr)^2 {{ M^2}\over{(M^2+Q^2)^2}}
         {2\over 3}\Big|\int d^2 l^{\prime} \bar\sigma_{(q \bar q)^{J=1}_T}
              (\vec l_\perp^{\prime 2},W^2) 
       F_T(Q^2, M^2, \vec l_\perp^{\prime 2}) \Big|^2,    
\label{(323)}
\ee
and
\be
\left. \frac{d\sigma_{\gamma^*_L p \rightarrow X^{J=1}_L p}}{dt dM^2} 
\right|_{t=0} = \Bigl({{c_0}\over{16\pi}}\Bigr)^2 {{Q^2}\over{(M^2+Q^2)^2}}
      {1\over 6}\Big| \int d^2 l^{\prime} \bar\sigma_{(q \bar q)^{J=1}_L}
         (\vec l_\perp^{\prime 2},W^2) 
      F_L(Q^2, M^2, \vec l_\perp^{\prime 2}) \Big|^2.  
\label{(324)}
\ee
They are related to (\ref{(317)}) and (\ref{(318)}) by the substitution rule
(\ref{(314)}) with subsequent integration over $z$.

In summary, solely based on the structure of the two-gluon-exchange dynamics 
of fig.~1, we derived the sum rules (\ref{(321)}) and (\ref{(322)}). They 
relate the transverse and longitudinal part of the total 
photoabsorption cross section to diffractive forward production of $q \bar q$ 
vector states, $X^{J=1}_{T,L} \equiv ( q \bar q)^{J=1}_{T,L}$. As the 
two-gluon-exchange dynamics from QCD for DIS at low $x$ can hardly
be doubted, the validity of GVD, or the 
generalized vector dominance/color-dipole picture (GVD/CDP) has thus been 
established. Conversely, a violation of the sum rules would imply a 
failure of the generic two-gluon exchange structure
\footnote{``Generic'' insofar, 
as the derivation is independent of the detailed specification of the lower 
vertices in fig.~1.}
that can hardly be imagined.

\section{The gluon structure function}

\setcounter{equation}{0}

\renewcommand{\theequation}{%
\mbox{\arabic{section}.\arabic{equation}}}

The gluon structure function is quantitatively related to the 
short-distance behaviour of the color-dipole
cross section. Comparing the expression for $\sigma_{\gamma^* p} (W^2 , Q^2)$ in 
terms of the color-dipole cross section, (\ref{(27)}), with the one based on 
the $\gamma^* g \rightarrow \gamma^* g$ interaction and the gluon-structure
function for $\vec r_\perp^2 \to 0$, one finds \cite{o,p}
\be
\sigma_{(q \bar q)p} (\vec r^{~2}_\bot , W^2) = \vec r^{~2}_\bot
\frac{\pi^2}{3} \alpha_s (Q^2)x g (x , Q^2).
\label{(401)}
\ee
According to (\ref{(32)}) to (\ref{(34)}), we have 
\be
   \sigma_{(q \bar q)p} (\vec r^{~2}_\bot ,W^2) =\vec r^{~2}_\bot
   \frac{\pi}{4} \int d \vec l^{~2}_\bot \vec l^{~2}_\bot
   \tilde\sigma_{(q \bar q)p} (\vec l^{~2}_\bot, W^2) . 
   \label{(402)}
\ee
Equating (\ref{(402)}) with (\ref{(401)}), we find
\be
   \alpha_s (Q^2)x g (x,Q^2) = \frac{3}{4\pi} \int d \vec l^{~2}_\bot 
   \vec l^{~2}_\bot \tilde\sigma_{(q \bar q)p} (\vec l^{~2}_\bot, W^2).
   \label{(403)}
\ee
Alternatively, the gluon structure function may be represented as    an
integral over $\vec l^{\prime 2}_\bot$,
\be
\alpha_s (Q^2)x g (x,Q^2) = \frac{1}{8\pi} \int d\vec l^{~\prime 2}_\bot
\vec l^{~\prime 2}_\bot\bar\sigma_{(q \bar q)^{J=1}_L } 
(\vec l^{~\prime 2}_\bot , W^2) .
\label{(404)}
\ee
Relation (\ref{(404)}) becomes identical to (\ref{(403)}) upon 
substitution of $\bar\sigma_{(q\bar q)_L^{J=1}}$ from (\ref{(12)}).
Relation (\ref{(404)}) says that the gluon structure function is 
related to the interaction of longitudinally polarized
$q\bar q$ vector states, $ (q\bar q)_L^{J=1} p \to (q\bar q)_L^{J=1} p$.
Combining (\ref{(404)}) with the short-distance expansion of (\ref{(40)}),
\be
\sigma_{(q \bar q)^{J=1}_L p} (\vec r^{~\prime}_\bot , W^2) =
\vec r^{~\prime 2}_\bot \frac{\pi}{4} \int d \vec l^{~\prime 2}_\bot
\vec l^{~\prime 2}_\bot \bar\sigma_{(q \bar q)^{J=1}_L p}
(\vec l^{~\prime 2}_\bot , W^2) ,
\label{(405)}
\ee
we find 
\be
\sigma_{(q \bar q)^{J=1}_L p} (\vec r^{~\prime 2}_\bot , W^2) = 
\vec r^{~\prime 2}_\bot \cdot 2 \pi^2 \alpha_s (Q^2) xg (x , Q^2) .
\label{(406)}
\ee

Relation (\ref{(406)}) is the analogue of (\ref{(401)}), if
the representation (\ref{(41)}) is used for $\sigma_{\gamma^*_{T,L}p}(W^2 , Q^2)$.
Relations (\ref{(404)}) and (\ref{(406)}) explicitly say that the 
gluon-structure function is determined by the dipole cross section for 
longitudinally polarized $q \bar q$ vector states. Contributions to the 
dipole cross section describing inelastic diffraction, $\gamma^*_{T,L} p 
\rightarrow (q \bar q)^{J\not= 1}_{T,L}p$, are irrelevant as far as the 
determination of the gluon structure function is concerned.

Finally, we may introduce the unintegrated gluon structure functions
\be 
     \alpha_s (Q^2)\tilde{{\cal F}}(x, \vec l^{~2}_\bot ) 
   = \frac{3}{4\pi}\vec l^{~2}_\bot \tilde\sigma_{(q \bar q)p} 
    (\vec l^{~2}_\bot , W^2)
\label{(407)}
\ee
as well as, in terms of $\bar\sigma_{(q \bar q)^{J=1}_L p}$, 
\be
     \alpha_s (Q^2) \bar{{\cal F}}(x, \vec l^{~\prime 2}_\bot) 
    = \frac{1}{8\pi}\vec l^{~\prime 2}_\bot
      \bar\sigma_{(q\bar q)^{J=1}_L p} (\vec l^{~\prime 2}_\bot, W^2).
\label{(408)}
\ee 
The gluon structure function is then given by
\be
     \alpha_s (Q^2) xg (x , Q^2) = \int d\vec l^{~2}_\bot 
     \alpha_s (Q^2)\tilde{{\cal F}} (x,\vec l^{~2}_\bot),
\label{(409)}
\ee 
and
\be
     \alpha_s (Q^2) xg(x , Q^2) = \int d\vec l^{~\prime 2}_\bot 
     \alpha_s (Q^2) \bar{{\cal F}} (x,\vec l^{~\prime 2}_\bot).
\label{(410)}
\ee

\section{The ansatz for the color-dipole cross section in the GVD/CDP}

\setcounter{equation}{0}

\renewcommand{\theequation}{%
\mbox{\arabic{section}.\arabic{equation}}}

In this section, we briefly remind the reader of the 
ansatz in the GVD/CDP that fits the experimental data for 
$\sigma_{\gamma^*_{T,L} p}(W^2 , Q^2)$. 
It was used in our previous derivation \cite{d} of the sum 
rules from section 3. 

We go back to the Fourier representation of the color-dipole cross section 
in (\ref{(39)}) in terms of the variable $\vec l^{~\prime}_\bot$ and to the 
cross sections for $(q \bar q)^{J=1}_{T,L}$ vector states in 
(\ref{(201)}) and (\ref{(202)}). 
In particular, let us assume the product
\be
     z(1-z)\tilde\sigma_{(q\bar q)p} (\vec l^{~\prime 2}_\bot z(1-z), W^2) 
   = f ( \vec l^{~\prime 2}_\bot , W^2)
\label{(501)}
\ee
to be independent of the variable $z$. In this case, the integrations over 
$dz$ in (\ref{(201)}) and (\ref{(202)}) can be carried out, and we have 
\be
     \bar\sigma_{(q \bar q)^{J=1}_T}(\vec l^{~\prime 2}_\bot, W^2) 
   = \bar\sigma_{(q \bar q)^{J=1}_L}(\vec l^{~\prime 2}_\bot, W^2)
   = f (\vec l^{~\prime 2}_\bot, W^2).
\label{(502)}
\ee
Substitution of a suitable function $f(\vec l^{~\prime 2}_\bot, W^2)$ into 
(\ref{(40)}) and (\ref{(41)}), as well as (\ref{(323)}) and (\ref{(324)}), 
yields predictions for 
$\sigma_{\gamma^*_{T,L} p} (W^2,Q^2)$ and for the diffractive production 
cross section for $\gamma^*_{T,L} p\rightarrow X^{J=1}_{T,L}p$ in the 
forward direction. 

Our ansatz \cite{e} in the GVD/CDP 
\begin{eqnarray}
     && z(1-z)\tilde\sigma (\vec l^2_\bot, W^2) = 
        z(1-z) \frac{\sigma^{(\infty)}}{\pi}
        \delta (\vec l^2_\bot - z(1-z) \Lambda^2 (W^2)) \nonumber \\
     && = \frac{\sigma^{(\infty)}}{\pi} \delta 
        (\vec l^{~\prime 2}_\bot - \Lambda^2 (W^2)) 
\label{(503)}
\end{eqnarray}
has precisely the form (\ref{(501)}), and according to (\ref{(502)}), it 
amounts to 
\be 
     \bar\sigma_{(q \bar q)^{J=1}_T} (\vec l^{~\prime 2}_\bot , W^2) = 
     \bar\sigma_{(q \bar q)^{J=1}_L} (
     \vec l^{~\prime 2}_\bot , W^2) = \sigma^{(\infty)} \cdot 
     \frac{1}{\pi} \delta (\vec l^{~\prime 2}_\bot - \Lambda^2 (W^2)) . 
\label{(504)}
\ee 
With respect to transverse position space, from (\ref{(32)}) or 
(\ref{(40)}), we find 
\be 
     \sigma_{(q \bar q)^{J=1}_T p} (\vec r^{~\prime}_\bot , W^2) =  
     \sigma_{(q \bar q)^{J=1}_L p}(\vec r^{~\prime}_\bot , W^2) = 
     \sigma^{(\infty)} (1 - J_0 ( r^\prime_\bot \cdot \Lambda (W^2)) . 
\label{(505)}
\ee 
As to the meaning of the ansatz (\ref{(503)}) to (\ref{(505)}),
it is worth noting that it is nothing else but an approximation of
the (unknown) distribution in the transverse gluon momentum
(multplied by $1/z(1-z)$), $\vec l_\bot^{~\prime 2}$, by a 
$\delta$-function that defines the effective value of 
$\vec l_\bot^{~\prime 2}$ via
\be
    \langle \vec l^{~\prime 2}_\bot \rangle_{W^2} 
  = \frac{\int d \vec l^{~\prime 2}_\bot
    \vec l^{~\prime 2}_\bot
    \bar\sigma_{(q\bar q)_L^{J=1}}
    \left(\vec l^{~\prime 2}_\bot,W^2 \right)}
    {\int d \vec l^{~2}_\bot \bar\sigma_{(q\bar q)_L^{J=1}}
    \left(\vec l^{~\prime 2}_\bot,W^2 \right)}
  = \Lambda^2(W^2).
\label{(506)}
\ee
The (expected) rise of the effective value of the transverse
momentum of the gluons is fitted to the experimental data.
In (\ref{(503)}) and (\ref{(505)}), $\Lambda^2 (W^2) = 
\Lambda^2 \left(\frac{Q^2}{x} \right)$ is a slowly increasing 
function of $W^2$ parameterized by a power law or a logarithm, 
\be
   \Lambda^2 (W^2) = \left\{ \matrix{&B ({{W^2}\over{W^2_0}}+1)^{C_2} , \cr
   &C^\prime_1 \ln \left(\frac{W^2}{W^{\prime 2}_0} + C^\prime_2 \right), \cr} 
   \right. 
\label{(507)}
\ee 
where $B, C_2, W^2_0$ and $C_1^\prime , C^\prime_2 , W^{\prime 2}_0$
are fit parameters and $\sigma^{(\infty)}$ is a cross section of 
typical hadronic size \cite{e}.  According to \cite{e},we have
$B=2.24\pm 0.43$ GeV$^2$, $W_0^2=1081\pm 124$ GeV$^2$ and
$C_2=0.27\pm0.01$. 

We finally give the gluon structure function for the ansatz (\ref{(504)}). 
It is given by 
\be
   \alpha_s (Q^2) xg (x,Q^2) = \frac{1}{8\pi^2} \sigma^{(\infty)} 
   \Lambda^2 (W^2) . 
\label{(508)}
\ee 
This is easily verified by substituting (\ref{(504)}) into 
(\ref{(404)}).

The ansatz (\ref{(503)}) with its special dependence on $z(1-z)$, according to 
(\ref{(504)}), is recognized as a specification of the elastic scattering 
of $q \bar q$ vector states. This is all that is needed to evaluate \cite{d}
the total cross sections (\ref{(312)}) and (\ref{(313)}) in momentum space, 
or (\ref{(41)}) in transverse position space. Substitution of (\ref{(504)})
into (\ref{(323)}) and (\ref{(324)})
yields \cite{d} the cross sections for diffractive production of  
$q \bar q$ vector states.

\setcounter{equation}{0}
\section{Duality, gluon structure function and saturation}
  It is worth emphasizing that the alternative approaches to a
theoretical description of DIS at low $x$ in terms of 
the color-dipole cross section
and in terms of the gluon structure function are to be 
considered as being dual to each other rather than excluding
each other.
Both descriptions rely on the two-gluon-exchange dynamical 
mechanism evaluated at low $x$.
While the GVD/CDP interprets the two-gluon-exchange
dynamics at low $x$ in terms of a $\gamma^*(q\bar q)$
transition with subsequent $(q\bar q)^{J=1} p \to (q\bar q)^{J=1} p$
scattering, the notion of the gluon-structure function relies on
$\gamma^* g \to \gamma^* g$ scattering.
The duality of the two pictures (in a restricted kinematical
domain)
becomes manifest in relations (\ref{(403)})
and (\ref{(404)}) that explicitly express the gluon structure 
function in terms of the (momentum-space expression for the)
color-dipole cross section.

\begin{figure}[htbp]\centering
\epsfysize=8cm
\centering{\epsffile{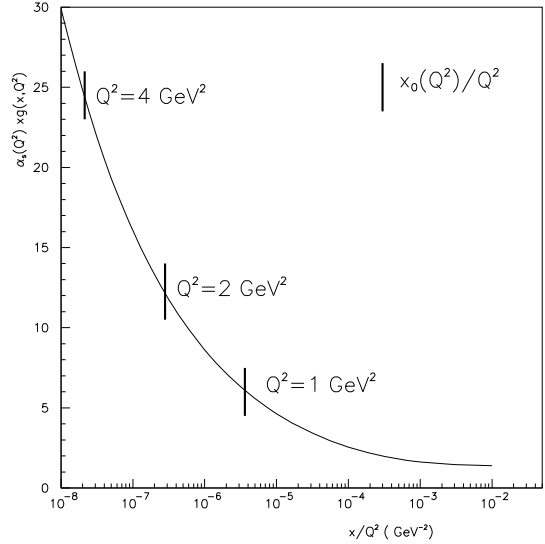}}
\caption{The GVD/CDP prediction of the gluon structure function,
$\alpha_s(Q^2)xg(x,Q^2)$,
as a function of a single variable $x/Q^2$. For $x\lsim
x_0(Q^2)$, as indicated, the conventional interpretation
of the gluon structure function breaks down and 
"saturation"sets in.}
\label{fig3}
\end{figure}
In fig.3, we show the gluon structure function  (\ref{(508)})
of the GVD/CDP.  It has the remarkable property of 
depending only on the energy $W$,
\bqa
       \alpha_s(Q^2)xg(x,Q^2) 
     &=& {1\over{8\pi^2}}\sigma^{(\infty)}
                               \Lambda^2(W^2) \nonumber\\
     &=& {1\over{8\pi^2}}\sigma^{(\infty)}
                               \Lambda^2({{Q^2}\over x}),
  \label{601}
\eqa
and accordingly, we plot it against
\be
    {1\over{W^2}} = {x\over{Q^2}}.  \label{602}
\ee
We urge experimentalists to plot the gluon structure function
( multiplied by $\alpha_s(Q^2)$) they extract from the 
measured cross sections against $x/Q^2$, in order to verify
the scaling behavior\footnote{Note that the scaling is to 
be considered as a LO QCD result that
is best fulfilled for $x\ll 0.1$.} of fig.3.
In fig.3,we also show selected values for the quantity $x_0(Q^2)$.  
The quantity $x_0(Q^2)$ defines 
the kinematical boundary of the region of $x\lsim x_0(Q^2)$
where saturation in the sense of \cite{e} 
\be
   \lim_{\tiny \matrix{ W^2\to \infty \cr Q^2 \to {\rm fixed}}}
    {{\sigma_{\gamma^*p}(W^2,Q^2)}\over
     {\sigma_{\gamma p}(W^2)}}  =1  \label{603}
\ee
sets in.
Saturation (\ref{603}) sets in for $\eta$ sufficiently small,
\be
    \eta \lsim \eta_0 \ll 1, \label{604}
\ee
where $\eta$ is the scaling variable,
\be
   \eta(W^2,Q^2) = {{Q^2+m_0^2}\over{\Lambda^2(W^2)}}, \label{605}
\ee
that determines the behavior of the total photoabsorption 
cross section\cite{e}
\be
   \sigma_{\gamma^* p}(\eta(W^2,Q^2)) ={{2\alpha}\over{3\pi}}
       \sigma^{(\infty)}\cases{
        \log(1/\eta), & for $\eta \ll 1$, \cr
        {1\over{2\eta}}, & for $\eta \gg 1$,}  \label{606}
\ee
and  $\Lambda^2(W^2)$ is given by (\ref{(507)}).
With (\ref{(507)}), the condition on $\eta$ in (6.4) is converted
into
\bqa
   x\lsim x_0(Q^2) &=& {{Q^2}\over{W_0^2}}
             {1\over{\Bigl( \bigl({{Q^2}\over{B\eta_0}}\bigr)^{1/C_2}
             -1\Bigr) }} \nonumber \\
       &\cong& {{(B\eta_0)^{1/C_2}}\over {W_0^2 (Q^2)^{1/C_2-1} }}.
 \label{607}
\eqa
The bound $x_0(Q^2)$ falls strongly with increasing $Q^2$,
since $x_0(Q^2)\cong 1/Q^4$.  The numerical values of $x_0(Q^2)/Q^2$
in fig.3 are based on $\eta_0=0.1$.

With respect to the interpretation of the transition to saturation,
it is useful\footnote{It is useful for the interpretation
of saturation, even though the simple scaling behavior in 
$\eta$ becomes a hidden one.}
to substitute the scaling variable (\ref{605}) into the cross
section (\ref{606}) and replace $\Lambda^2(W^2)$ by the gluon
structure function (\ref{(507)}).  We obtain,
\be
    \sigma_{\gamma^* p}(\eta({{Q^2}\over x},Q^2)) 
   = {{2\alpha}\over{3\pi}}\sigma^{(\infty)}
    \cases{ \log{{8\pi^2\alpha_s(Q^2)xg(x,Q^2)}\over
                 {\sigma^{(\infty)}(Q^2+m_0^2)}}, &
                         $\eta \ll 1$, \cr
         {{8\pi^2\alpha_s(Q^2)xg(x,Q^2)}\over
                 {2\sigma^{(\infty)}Q^2}}, &
                         $\eta \gg 1$.   }  \label{608}
\ee
According to (\ref{608}), the transition to the saturation
region (obviously) does not imply that the gluon-structure
function ceases to increase with decreasing $x$.  
The onset of saturation depends 
on $Q^2$ via $x_0(Q^2)$, and it simply
means that the approximation of $\eta\gg 1$ in (\ref{606})
and (\ref{608}) breaks down, and the logarithmic
(soft) behavior sets in.  Note that the region of
$\eta \gg 1$ is the one where the logarithmic derivative of the
structure function $F_2=(Q^2/4\pi^2\alpha) 
\sigma_{\gamma^*p}(W^2,Q^2)$ yields the gluon structure function
\cite{e} as a result of the evolution equations.

To summarize, saturation does not mean that the gluon structure
function ceases to rise.  The gluon structure function rises 
indefinitely for $1/W^2=x/Q^2\to 0$.  At any fixed $Q^2$,
however, for $x\lsim x_0(Q^2)$, the conventional connection between
$F_2$ and the gluon density breaks down,
and saturation in the sense of (\ref{603}) sets in.
Alternatively, for any fixed $x$, however small,
the conventional evolution takes place provided $Q^2$ is 
sufficiently large.

\section{Conclusion}

We end with a brief summary:

i) Quantum chromodynamics, in particular  the generic
two-gluon-exchange, valid at low $x$, implies a 
representation of the total photoabsorption cross 
section as a sum over the mass spectra of diffractive production 
of $(q\bar q)$ vector states.  
In this sense, the generalized vector dominance picture is a
consequence of QCD.  
The GVD/CDP may thus be considered established insofar as its violation
would falsify the underlying generic two-gluon-exchange
structure  - an assumption hardly 
questionable from all we know about quark and antiquark interactions in QCD.

ii) The kinematic domain of the GVD/CDP, $x\lsim 0.1$ and 
$Q^2$ arbitrary, including $Q^2=0$, allows one to estimate 
the kinematic domain where the GVD/CDP and the description in terms of
the gluon structure function are dual to each other.
Apart from the usual restriction of $Q^2\gsim Q_0^2>0$, we find
that the duality domain is bounded by $x\gsim x_0(Q^2)$, where
$x_0(Q^2)$ is exceedingly small and decreases strongly with
increasing $Q^2$.  For $x\lsim x_0(Q^2)$ saturation sets in.

\end{document}